\documentclass{article}

\usepackage{PRIMEarxiv}

\usepackage[utf8]{inputenc} 
\usepackage[T1]{fontenc}    
\usepackage{hyperref}       
\usepackage{url}            
\usepackage{booktabs}       
\usepackage{amsfonts}       
\usepackage{nicefrac}       
\usepackage{microtype}      
\usepackage{lipsum}
\usepackage{fancyhdr}       
\usepackage{graphicx}       
\graphicspath{{media/}}     

\usepackage{amssymb}
\usepackage{amsmath}
\usepackage{float}
\usepackage{soul}
\usepackage{lineno}
\usepackage[dvipsnames]{xcolor}

\title{Super-resolved Second Harmonic Generation Imaging by Coherent Image Scanning Microscopy
}

\author{
  Dekel Raanan \\
  Department of Physics of Complex Systems \\
  The Weizmann Institute of Science  \\
  Rehovot 76100, Israel\\
  \texttt{dekelr@gmail.com} \\
   \And
  Man Suk Song \\
  Department of Condensed Matter Physics\\
  The Weizmann Institute of Science \\
  Rehovot 76100, Israel\\
  \texttt{man-suk.song@weizmann.ac.il} \\
   \And
  William A Tisdale \\
  Department of Chemical Engineering\\
  Massachusetts Institute of Technology \\
  Cambridge, Massachusetts 02139, United States\\
  \texttt{tisdale@mit.edu} \\
   \And
  Dan Oron \\
  Department of Molecular Chemistry and Materials Science\\
  The Weizmann Institute of Science\\
  Rehovot 76100, Israel  \\
  \texttt{dan.oron@weizmann.ac.il} \\
}

\begin{document}
\maketitle

\begin{abstract}
We extend image scanning microscopy to second harmonic generation (SHG) by extracting the complex field amplitude of the second-harmonic beam. While the theory behind coherent image scanning microscopy (ISM) is known, an experimental demonstration wasn't yet established. The main reason is that the naive intensity-reassignment procedure cannot be used for coherent scattering as the point spread function is now defined for the field amplitude rather than for the intensity. We use an inline interferometer to demonstrate super-resolved phase-sensitive SHG microscopy by applying the ISM reassignment machinery on the resolved field. This scheme can be easily extended to third harmonic generation and stimulated Raman microscopy schemes. 
\end{abstract}

\keywords{Coherent microscopy \and Image scanning microscopy \and Super resolution}

\section{Introduction}
Far field optical super-resolution employing incoherent signals, most commonly fluorescence, is today a reality, where in some cases the resolution almost reaches the ultimate scale of the size of a molecule \cite{Gwosch2020}. Several superresolution modalities (such as PALM and STORM) rely on turning individual emitters "on" and "off", thus allowing to perform localization of dilute sub-ensembles of emitters \cite{Rust2006,Betzig2006}. Other techniques (such as STED or SAX) employ saturation or high order non-linearities \cite{Hell1994,Fujita2007}. The use of correlations of fluorescence fluctuations, classical or quantum, has also been shown to enable surpassing the diffraction limit \cite{Dertinger2009,Schwartz2013}. A different approach to perform super-resolution microscopy relies on tailoring the spatial excitation profile of the excitation beam. Typically these methods achieve moderate resolution enhancement as compared to the previously described methods, yet they are more general and often experimentally simpler. Two well-established examples of such schemes are structured illumination microscopy (SIM) \cite{Gustafsson2000,Futia2011} and image scanning microscopy (ISM) \cite{SHEPPARD1988,Mueller2010}. Both schemes achieve a resolution enhancement of up to a factor of two as compared to wide-field illumination. 

Applying super-resolution techniques to coherent signals is a difficult task for several reasons. First, unlike resonant electronic absorption, standard non-resonant coherent processes such as second harmonic generation (SHG) and third-harmonic generation (THG), cannot typically be turned on and off. Thus, localization of such samples can only be performed if the non-linear sources are sparse enough to begin with, and cannot be made on dense samples. Second, saturating resonant non-linear processes (such as stimulated Raman scattering and coherent anti-Stokes Raman scattering), even though previously demonstrated \cite{Gong2019,Choi2018,Kim2017}, requires extremely high optical power densities which are often incompatible with microscopy applications. Some reports on implementation of SIM and related techniques to coherent imaging exist \cite{Field2016,Yeh2018}, yet do not resolve the field amplitude and as such must involve a component of phase retrieval which often interferes with the image reconstruction process. Still, due to the limited possibilities to tweak the properties of the non-linear emitter, the favored candidate for performing super resolution in coherent non-linear scattering, and particularly for non-resonant processes, is by exploiting the spatial properties of the excitation beam.  

Image scanning microscopy, which was invented in the late 1980s \cite{SHEPPARD1988} and experimentally demonstrated in 2010 by Muller and Enderlein \cite{Mueller2010} is based on the same setup as the confocal microscope invented by Minsky in 1957 but gives access to the full spatial resolution afforded by a closed pinhole while integrating over the entire signal as measured by a large detector. This is enabled by the use of a pixelated detector instead of a bucket detector, where each pixel acts as a small pinhole, and is followed by a simple post-processing stage known as pixel reassignment. The origin of the reassignment machinery is that the detection point spread function (PSF) is shifted for the different detectors due to parallax. As a consequence of this shift, the different pixels retain a shifted image as compared to an image obtained from a detector placed on the optical axis. It is important to note, however, that the simple reassignment process only works well if the PSF is well-defined and shift-invariant \cite{Rossman2021}. For incoherent processes such as fluorescence or spontaneous Raman scattering the PSF is defined for both the field and the intensity, as the field from different emitters is not phase-correlated. Thus, for incoherent sources pixel-reassignment is done directly on the raster-scanned images obtained by the different pixels of a camera. In contrast, when considering coherent signals, the PSF is only defined for the complex field amplitude. In this case the intensity image consists of interference between different sources and hence the PSF is ill-defined for intensity and reassignment cannot be directly applied to it as was previously attempted \cite{Gregor2017}. Instead, the reassignment process should be done on the resolved field. In this work we rigorously perform coherent ISM on two types of SHG objects, including an amplitude object and a phase object, and compare the performance to an open-aperture confocal measurement in terms of both resolution and contrast.

Before proceeding to the experimental demonstration let us first introduce a simple calculation based on Gaussian beams to assess the expected resolution increase and the necessary pixel reassignment process for the case of SHG. We assume a Gaussian envelope for both the excitation and detection PSFs having extents $\sigma_{1\omega}^{exc}$ and $\sigma_{2\omega}^{det}$, respectively. These depend on both the numerical aperture and the wavelength. As harmonic generation involves the creation of frequencies which are  separated by an octave, it is important to consider the difference between the excitation and detection wavelengths which is much larger than the stokes shift typically associated with fluorescence or spontaneous Raman scattering. Moreover, in coherent scattering phase matching considerations have to be taken into account. Thus, the simplest description is of a forward scattered signal, as also demonstrated experimentally below. We note that this can be easily generalized to epi-detection. Assuming that the excitation beam defines the optical axis, the excitation PSF is:

\begin{equation} 
\label{ExcPSF}
h_{SHG}^{exc}=\left[e^{-\left(\frac{R_{source}}{{\sigma_{1\omega}^{exc}}}\right)^2}\right]^2
\end{equation}

and a corresponding detection PSF, centered around the source:
\begin{equation} 
\label{DetPSF}
h_{SHG}^{det}=e^{-\left(\frac{R_{det}-R_{source}}{\sigma_{2\omega}^{det}}\right)^2}
\end{equation}

\begin{figure}[H]
\centering\includegraphics[width=8cm]{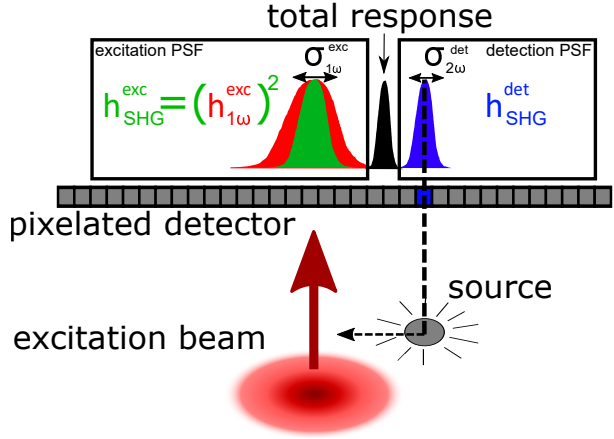}
\caption{Excitation and detection PSFs for SHG-ISM. The excitation beam is centered around the optical axis, and for SHG, is squared to obtain the SHG excitation PSF. The detection PSF is centered around the individual pixel and represents the response to different locations of the source, assuming flat illumination.}
\label{CISM_Reassgn}
\end{figure}

$R_{source}$ and $R_{det}$ represent the location of the emitter and a specific detector in the array with respect to the optical axis on the sample plane. The total PSF, which is the multiplication of the excitation and detection PSFs is illustrated in figure \ref{CISM_Reassgn} in black together with the excitation PSF in green (being the square of the excitation field, denoted in red) and the detection PSF in blue. We find that the total PSF, similarly to the incoherent ISM case, is linearly shifted and narrowed as compared to both the detection and excitation PSFs, nonetheless relates to the field amplitude rather than to the intensity. Furthermore, since the phase-front is known, quantifying the deviation of the reassignment shift as a function of the pixel number and comparing it to the theoretical shift allows to compensate for defocusing. The reassignment shift is found by calculating the location of the source with respect to the optical axis that provides the maximum signal at $R_{det}$:

\begin{equation} 
\label{ReassgnVal}
R_{shift}=R_{det}\cdot\frac{(\sigma_{1\omega}^{exc})^2}{(\sigma_{1\omega}^{exc})^2+2(\sigma_{2\omega}^{det})^2}
\end{equation}

This yields the procedure for obtaining the SHG image from the measurements. The total electric field at the image plane is obtained by convolving the total PSF $h=h^{exc}*h^{det}$ with the scattered field in the object plane $E_{obj}$. Here, since the signal is coherent, the convolution must be done using the field PSF, rather than the intensity PSF:
\begin{equation} 
\label{SignlConv}
E_{image}={E_{obj}\circledast h}
\end{equation}

for which the estimated resolution follows:
\begin{equation} 
\label{ReassgnRes}
\sigma_{T}=\frac{\sigma_{1\omega}^{exc}\cdot\sigma_{2\omega}^{det}}{\sqrt{(\sigma_{1\omega}^{exc})^2+2(\sigma_{2\omega}^{det})^2}}
\end{equation}


We note that to obtain the reconstructed SHG intensity image one must take the absolute value squared of the obtained field.

\section{Experiment}
To reconstruct the SHG electric field we need to perform a phase-sensitive measurement for which a reference SHG beam is required. This is usually done using an interferometrically stabilized SHG reference, which introduces additional complexity to the optical setup \cite{Gao2018,Yazdanfar2004}. In the experiment described below, we chose to use a reference beam co-propagating with the fundamental beam which greatly simplifies the phase stabilization \cite{Bancelin2016}. 

\begin{figure}[H]
\centering\includegraphics[width=13cm]{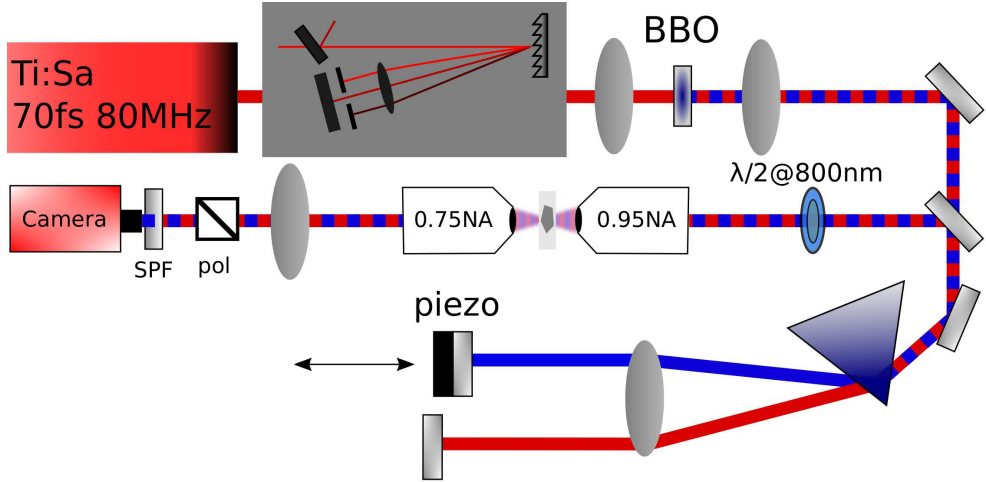}
\caption{Experimental setup. The laser is followed by a compressor and a BBO crystal which creates the reference SHG beam that is needed for retrieving the SHG field from the sample. The beams are then sent into a reflective prism-based 4f pulse shaper which both matches the group delay and scans the relative phase between the signal and reference beams. The beams are then directed into the scanning microscope which is followed by a polarizer and a short-pass filter to isolate the SHG signal.}
\label{ExperimentalSetup}
\end{figure}

Our experimental setup for realizing coherent SHG ISM is described schematically in figure \ref{ExperimentalSetup}. Our laser source is a 70fs Ti:Sapphire ultrafast oscillator (MaiTai DeepSee, Spectra Physics) which is first sent into a home-built pulse compressor where a narrow (circa 1nm) spectral window is selected to avoid high order dispersion effects. The fundamental laser beam is then focused on a type-I BBO crystal to create an orthogonally polarized 2$\omega$ reference beam. To facilitate phase shifting as well as provide simple control of the relative delay and dispersion of the fundamental and reference beams we chose to send both co-propagating beams into a prism-based 4f shaper which both matches the group delay between the two harmonics and scans the relative phase between the beams using a piezoelectric transducer placed in the blue arm. Finally the two beams are recombined and focused onto the sample by an apochromatic objective lens (Nikon, Apochromat 60X 0.95NA). The sample is raster scanned by a nano positioning stage (MCL, Nano-3D200) and the scattered SHG signal is collected by a second apochromat objective (20X 0.75NA) which is followed by a thin film polarizer to adjust the polarization state of the reference and signal beams. The interference between the 100X magnified SHG signal and reference beams is detected by an EMCCD camera (iXon 3, Andor). 
The pixel dwell time is of the order of 150ms, and consists of the time required to acquire four images, corresponding to four phase steps implemented by the piezo phase-scanner to retrieve the amplitude and phase of the SHG electric field (see supplementary information for details). The imaging process is synchronized by a computer controlled digital to analog card.

\begin{figure}[H]
\centering\includegraphics[width=13cm]{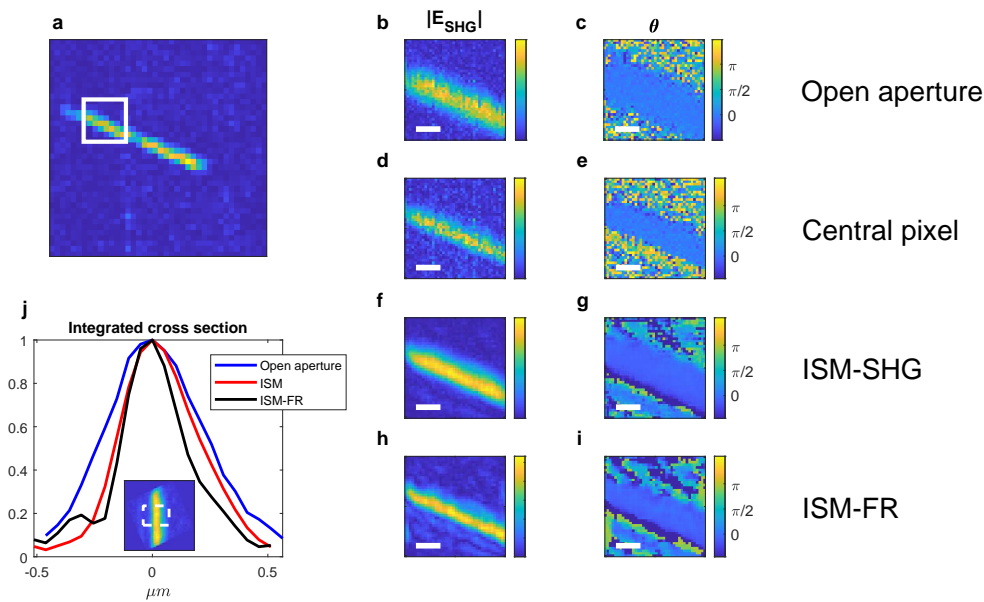}
\caption{An SHG image of a nanowire. (a) large field of view image of the rod. The rectangle marks the region which is imaged in (b-i). The field amplitude and phase are extracted and plotted to the right by considering (b-c) direct integration over the camera pixels, (d-e) the central pixel only of the camera, (f-g) the SHG-ISM image obtained after applying the reassignnment machinery on the field images as resolved by the individual camera pixels, and (h-i) is the SHG-ISM image followed by a Fourier reweighting analysis. (j) shows the cross-section of the nanowire for open aperture, ISM, and ISM-FR. The regime which is integrated is the one marked by the white dashed rectangle in the inset. Scale bar: 500nm.}
\label{NanoRod}
\end{figure}

We start by observing a simple amplitude object which can assist in determining directly the resolution improvement. Our sample is an InAs nanowire with a length of a few microns and a width of circa 50 nanometers, thus allowing us to estimate the PSF along the narrow direction. A large field-of-view image of the nanowire is presented in figure \ref{NanoRod} (a), which is followed by a smaller field of view, high resolution measurement to the right. We first mimic an open-aperture confocal microscope by integrating the signal detected by all the camera pixels. The field reconstruction is then performed by treating the camera as a single bucket detector (b-c). We then plot the resolved field amplitude as obtained from the central pixel of the camera only, ignoring the contribution of all the other pixels (d-e). This provides a lower SNR as compared to the open aperture measurement as less signal is collected, but a higher resolution as expected from a closed aperture detection. Next we combine the contribution of all the different camera pixels, taking into account the parallax shift by reassignment of the field (f-g). This results in a similar resolution as the single-pixel measurement, yet with the advantage of exploiting all the signal that is collected by the camera. Then one may use Fourier reweighting on the ISM image to compensate for the reduction of the response at high spatial frequencies in the optical transfer function (h-i). This is done by a Gaussian approximation of the PSF and a corresponding normalization of the image spatial frequency spectrum by the optical transfer function using:

\begin{equation} 
\label{FourierReweightEq}
E_{FR} = F^{-1}\left[\frac{F\left[E_{ISM}\right]}{OTF+\epsilon\cdot\frac{f_r}{f_c}}\right]
\end{equation}

$F$ and $F^{-1}$ denote the fourier transform and the inverse fourier transform, respectively. $E_{FR}$ is the extracted field after fourier reweighting, $E_{ISM}$ is the detected ISM field. OTF is the optical transfer function which is the fourier transform of the PSF. $\epsilon$ is typically set to a few times the noise floor of the OTF, $f_{r}$ is the magnitude of the spatial frequency, and $f_{c}$ is the cutoff spatial frequency, set to be the approximate spatial frequency where the OTF reaches $\epsilon$.
To estimate the resolution of each of the analysis techniques we plot the cross-section perpendicular to the wire in figure \ref{NanoRod} (j). The curves are obtained by integrating the signal along the wire in the region that is plotted in the white dashed rectangle in the inset of (j). The FWHM for the three curves are 500nm 390nm and 290nm for open aperture, ISM, and ISM-FR respectively. We note that the observed resolution is lower than the maximal theoretical resolution (FWHM) as we do not entirely fill the back aperture of the excitation objective. Furthermore, we signify that the resolution presented here is for the field amplitude, while the resolution for the intensity is approximately 1.4x higher.

\begin{figure}[H]
\centering\includegraphics[width=13cm]{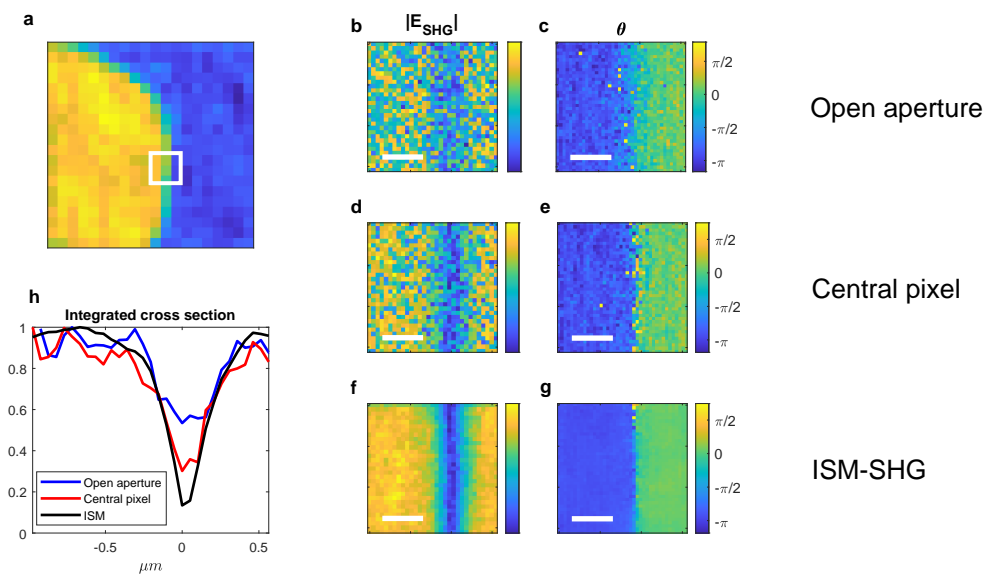}
\caption{(a) Large field of view image of the right part of the hexagon domain. The field of view in (b-g) is marked by the white rectangle. (b-g) are a comparison between (b-c) open-aperture confocal microscopy implemented by direct integration of the signal captured by the camera, (d-e) a point detector implemented by using only a single camera pixel and (f-g) SHG-ISM, obtained by reassigning the field images resolved by the different pixels of the camera to overlap their field of view. The cross-section along the interface, obtained by vertical integration and normalization is plotted in (h). The curves manifest the superiority of the ISM image over the image obtained by a direct integration of the signal collected by the camera. Scale bar: 500nm}
\label{ComparisonRtheta}
\end{figure}

Phase objects are typically more challenging to image. To show the utility of coherent ISM in this context we choose a sample made of Stoichiometric $LiTaO_3$ with hexagon-shaped domains created by electric field poling. Due to the different poling direction, there is a relative $\pi$ phase shift in the generated SHG signal inside and outside the domains. We first examine the resolution obtained by integrating the signal from the camera, as if the camera were replaced with a bucket detector. The extracted field amplitude and phase, presented in figure \ref{ComparisonRtheta} (b-c), clearly show a $\pi$ phase-step at the boundary between the two sides of the interface. Comparing the reconstructed image obtained by the central pixel of the camera in figure \ref{ComparisonRtheta} (d-e) to the open aperture image reveals a narrower, deeper feature of the interface. The reason for this is the parallax demonstrated in figure (S2). Each camera pixel captures a slightly translated image as to a slightly different field of view. Summing all the images altogether results in a widened central feature with a lower contrast as compared to the image obtained by a single pixel, or to a reconstruction obtained by reassigning the individual images to overlap their field of view, as plotted next in (f-g). This reduction of contrast was also discussed in the literature by Wilson and Sheppard \cite{Wilson1984} 
for closed-aperture confocal microscopy.
Figure \ref{ComparisonRtheta} (h) shows the normalized vertical integration of the images presented in figure \ref{ComparisonRtheta} (b,d,f) and demonstrates the previous claims about the width and contrast of the interface.


\section{Summary}
To summarize, coherent super-resolved microscopy could be used as a complementary tool to incoherent super-resolved imaging, gaining from both the natural advantages of non-linear processes such as a narrower PSF and sample specificity, in addition to the extra resolution obtained by the super-resolution method used. In this work we assisted the spatial profile of the excitation beam and a phase-stable inline interferometer to gain both the resolution enhancement of ISM and to reconstruct the amplitude and phase of the emerging SHG field. This allowed us to apply the pixel-reassignment machinery and to fully benefit the detection PSF. We found that the resolution enhancement and reassignnment process are conceptually identical to the more common incoherent case, but the analysis should be done on the extracted field rather than on the measured intensity. The inherent advantages of non linear interactions in addition to the generality of this scheme could play a significant role in obtaining even higher resolution and larger penetration depth in 3rd order coherent processes such as coherent Raman scattering and third harmonic generation, which are already extensively used by the biological community.

\section*{Acknowledgments}
The authors thank Prof. Ady Arie for providing SHG phase objects and Dr. Hadas Shtrikman for providing InAs nanowires. In addition the authors thank Uri Rossman for fruitful discussions.

\section{Disclosures}
The authors declare no conflicts of interest.

\section{Funding}
The authors gratefully acknowledge funding by the Israeli science foundation and the Crown center of photonics. DO is the incumbent of the Harry Weinrebe professorial chair of laser physics. WAT was supported by the Lin Family Faculty Research Innovation Fund.

\bibliographystyle{unsrt}  
\bibliography{ms}

\end{document}


\maketitle


\section{Supplementary}
In this supporting information we explain the reconstruction procedure of the fields and demonstrate the necessity of the reassignment procedure to re-gain the extra resolution provided by the detection PSF. We capture the interference pattern between the signal and a references beam at four different phases. These measurements are used to retrieve the amplitude and phase of the SHG beam. We remove the background of the camera in addition to a constant phase front which we relate to the reference beam. Exemplary images at four different phases as detected by the camera are presented in figure (S1). 

\begin{figure}[H]
\centering\includegraphics[width=8cm]{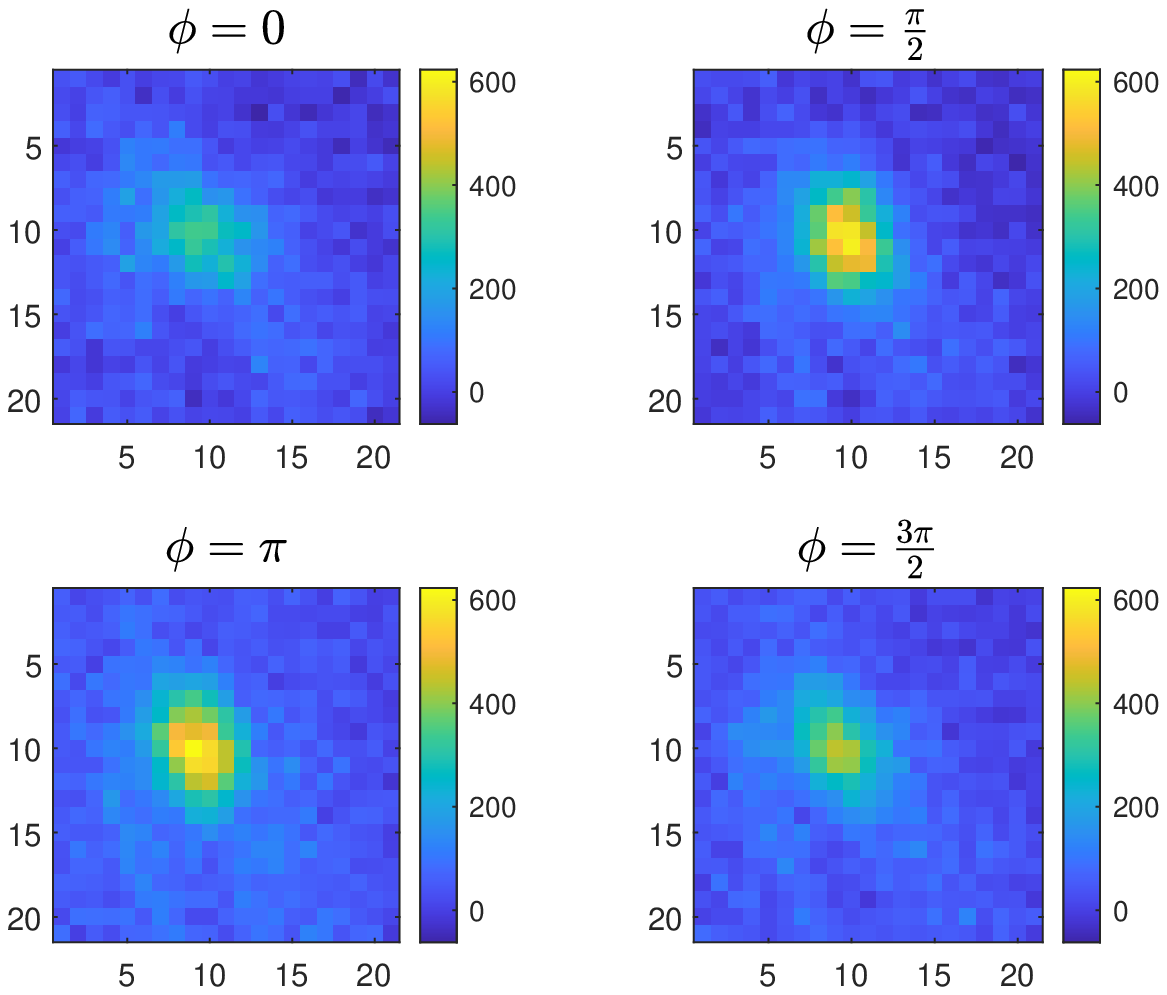}
\caption{Raw images from camera at a specific sample position at four different relative phases: 0,$\pi/2$,$\pi$,$3\pi/2$. From these we retrieve the phase and amplitude of both the reference and signal fields.}
\label{RawCameraImages}
\end{figure}

The interference signal results in the following intensity pattern on the camera:
\begin{equation} 
\label{Intensity}
I\left(\phi \right)=|E_{1}|^2+|E_{2}|^2+2\eta|E_{1}||E_{2}|\cos{\left(\theta+\phi\right)}
\end{equation}

where $E_{1}$ and $E_{2}$ are the reference and signal fields at $2\omega$, $\theta$ is the relative phase between the fields and $\phi$ is an externally controlled phase which we alter between 0,$\frac{\pi}{2}$,$\pi$,$\frac{3\pi}{2}$. $\eta$ is the interference contrast which is smaller than 1 as to imperfect spectral overlap between the signal and reference pulses. We then find that:

\begin{equation} 
\label{Theta}
\theta=atan\left(-\frac{I\left(\frac{\pi}{2}\right)-I\left(\frac{3\pi}{2}\right)}{I\left(0\right)-I\left(\pi\right)}\right)
\end{equation}

and the reconstructed field amplitudes are:
\begin{equation} 
\label{FieldAmp}
\left|E\right|_{1,2}^{2}=\frac{\left(I\left(0\right)+I\left(\pi\right)\right)\pm\sqrt{\left(I\left(0\right)+I\left(\pi\right)\right)^{2}-\frac{1}{\eta^{2}}\left(\left(I\left(0\right)-I\left(\pi\right)\right)^{2}+\left(I\left(\frac{\pi}{2}\right)-I\left(\frac{3\pi}{2}\right)\right)^{2}\right)}}{4}
\end{equation}

Figure (S2) demonstrates the shifted images obtained by the different camera pixels due to parallax. We plot the real part of the electric field, which flips its sign around the boundary between the inversely-poled domains. The location of the boundary between the domains linearly drifts with the distance of the pixel from the optical axis.

\begin{figure}[h]
\centering\includegraphics[width=6cm]{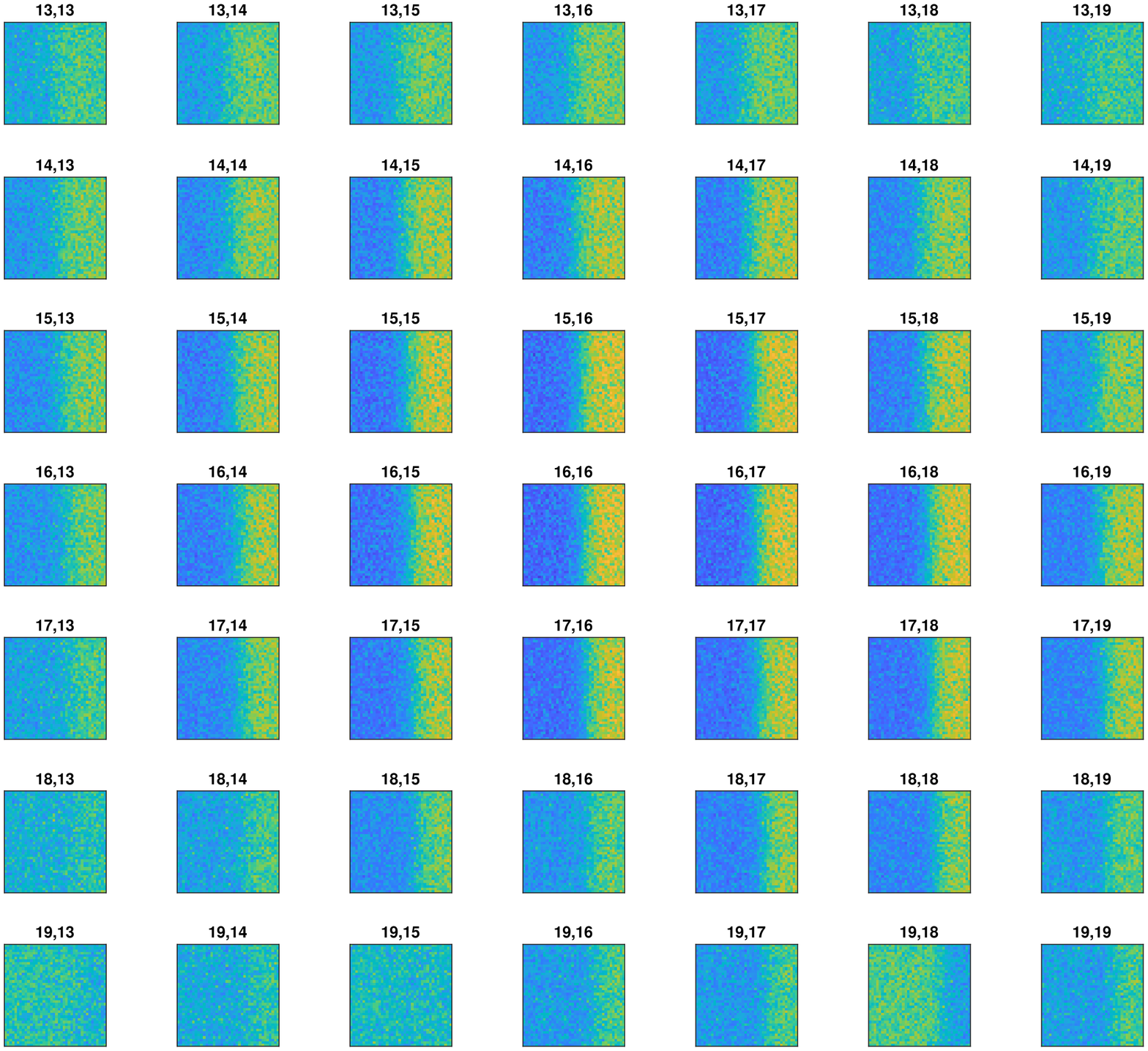}
\caption{Parallax images obtained by the different pixels of the camera. The camera pixel is mentioned above the corresponding image.}
\label{ParalxFig}
\end{figure}